# Cosmological bouncing solutions and their stability in teleparallel gravity

Adnan Malik,[1,2,3,4,*] Aimen Rauf,[3,†] Kazuharu Bamba,[5,‡] Wenbin Lin,[1,2,§] and Fatemah Mofarreh[6,¶]

[1]*School of Nuclear Science and Technology, University of South China, Hengyang 421001, China*
[2]*School of Mathematics and Physics, University of South China, Hengyang, 421001, China.*
[3]*Department of Mathematics, University of Management and Technology, Sialkot Campus,Pakistan.*
[4]*Center for Theoretical Physics, Khazar University, 41 Mehseti Str., AZ1096 Baku, Azerbaijan*
[5]*Faculty of Symbiotic Systems Science, Fukushima University, Fukushima 960-1296, Japan*
[6]*Mathematical Science Department, Faculty of Science,*
*Princess Nourah Bint Abdulrahman University, Riyadh 11546, Saudi Arabia*

**Abstract**

The cosmological dynamics in the early universe are investigated to explore the possibility of the sign reversal of the Hubble parameter as a key feature of non-singular bouncing cosmological solutions in higher-order torsion gravity. The self-consistent multiple cosmological regimes are studied, such as the accelerated expansion, ultra-relativistic, radiation-dominated, sub-relativistic, dust, and stiff matter phases, for three distinct parametrizations of the scale factor: power-law, exponential, and hybrid forms. In particular, five characteristic bouncing scenarios are analyzed: symmetric bounce, super-bounce, oscillatory bounce, matter bounce, and Type IV singularity-free bounce, so that the gravitational Lagrangian can be reconstructed to satisfy bounce conditions at the bounce time. It is found that each scenario requires a violation of the null energy condition, implying the presence of exotic matter with an effective equation of state to drive both the bounce and late-time cosmic acceleration. As a result, it is explicitly demonstrated that higher-order torsion gravity naturally incorporates the bouncing solutions without introducing ad hoc matter fields, providing a possible geometric framework for non-singular early universe evolution. Furthermore, the consistency of the bouncing solutions with the observational constraints of the cosmic microwave background and gravitational wave spectrum is shown, while offering testable predictions for primordial perturbations.

## I. INTRODUCTION

Two main approaches have been proposed to explain the two distinct phases of accelerated expansion the universe has gone through- one in its early stages and another occurring later. One key approach involves studying modified gravity theories. The idea behind modified gravity is to extend or adjust Einstein's theory of general relativity, making it more compatible with quantum gravity and explaining the observed accelerated expansion [1, 2]. However, finding the right formulation of modified gravity remains a challenge. Many studies begin with the standard curvature-based framework and then modify or extend the Einstein-Hilbert action. One common example is the $f(R)$ theory, where the Lagrangian becomes a nonlinear function of the curvature scalar. These modifications can lead to new cosmic behaviors, such as non-standard singularities, where the universe's scale factor, Hubble parameter, and energy density stay constant. Still, the rate of change of the Hubble parameter ($\dot{H}$) becomes infinite [3]. This type of singularity is problematic because it prevents the universe from evolving further. The long-term evolution of cosmological models must consider possible finite-time future singularities, which have been comprehensively reviewed in [4]. Our model's avoidance of such singularities represents an important advantage over many alternative dark energy scenarios. Astrophysical and cosmological observations are crucial in improving our understanding of the fundamental components of the universe. For example, studying anisotropies in the Cosmic Microwave Background (CMB) reveals valuable insights about the early universe. High-precision CMB measurements from satellites like COBE, WMAP, and Planck have shown that baryonic matter constitutes only a small fraction of the energy content of the universe [5–11]. Additionally, observations of galaxy rotation curves [12], galaxy clustering [13], and X-ray emissions [14] suggest that about 26% of the universe's total energy is in the form of dark matter. Dark matter plays a crucial role in forming galaxy clusters and large-scale structures, as baryonic matter alone cannot account for their presence, especially at high redshifts. Recent observations of type Ia supernovae [15, 16], combined with data from CMB [17–20], have revealed that the universe's expansion shifted from slowing down to speeding up around 7-8 billion years ago, corresponding to a redshift of about 0.6-0.8 [21]. This discovery led to the introduction of dark energy, a mysterious form of energy that is believed to make up about 69% of the total energy of the universe, which drives this late-time acceleration. Dark energy cannot be explained by known matter or energy fields. To explain this acceleration, many studies have explored modified gravity theories as a potential solution, proposing extensions to general relativity [22–34].

[*]Electronic address: adnan.malik@zjnu.edu.cn;adnan.malik@skt.umt.edu.pk;adnanmalik_chheena@yahoo.com
[†]Electronic address: aimenrauf90@gmail.com
[‡]Electronic address: bamba@sss.fukushima-u.ac.jp
[§]Electronic address: lwb@usc.edu.cn
[¶]Electronic address: fyalmofarrah@pnu.edu.sa

Another approach that has gained attention recently is the idea of cosmological bouncing models. These models offer a way to avoid the Big Bang singularity problem by suggesting that the universe first contracts to a minimum size and then bounces back, expanding once again [35–40]. Such models may also provide information on potential quantum gravity effects in the early universe [41–43]. Bouncing cosmology has emerged as a promising alternative to the standard inflationary paradigm, showing the capability to generate a scale-invariant power spectrum similar to that produced by inflationary models [44, 45], particularly in scenarios like the matter bounce [46]. These concepts are explored in various references [47, 48]. Another intriguing concept is the wormhole, a hypothetical bridge connecting two different universes or distant regions within the same universe. If a wormhole permits travel in both directions, it is termed a traversable wormhole [49, 50]. For traversable wormholes to exist, exotic matter that violates the null energy condition (NEC) is required [51]. This necessity has led researchers to explore various solutions, including dynamical wormholes [52], brane wormholes [53], and generalized chaplygin gas [54], all aimed at reducing the extent to which the NEC is violated. Modified gravity theories have also been considered as potential frameworks that could support wormhole structures [55]. Researchers have identified physically plausible wormhole solutions that comply with the NEC for isotropic and barotropic cases. However, the NEC is violated in anisotropic wormholes within generalized teleparallel gravity. Non-commutative geometry has also been investigated for potential wormhole solutions, revealing asymptotically flat and non-flat solutions in four and five dimensions, respectively. In the cosmological field, there is growing interest in different types of modified gravity theories, such as $f(R)$ gravity with $R$ the scalar curvature, $f(T)$ gravity with $T$ the torsion scalar, $f(G)$ gravity with $G$ the Gauss-Bonnet invariant and higher-order curvature gravity with non-minimally matter coupling, as well as various dark energy models [56–62]. Other studies have focused on testing the models $f(Q)$ gravity with $Q$ the non-metricity scalar [63–67]. Bounce cosmologies have been extensively studied in various modified gravity frameworks. Notable examples include string-inspired bounce scenarios [68] and Galileon-based bounce models (G-bounce) [69], each offering different mechanisms to avoid the initial singularity while maintaining consistency with cosmological observations. Some bouncing cosmological models through cosmic evolution with cyclical universe model, and non-singular cosmology have been investigated in [70–73].

Building on these findings, this study aims to investigate how effective $f(T)$ gravity is in describing entropy-corrected density scenarios in both power-law and logarithmic forms as alternative dark energy models. Our research delves into the intriguing realm of higher-order torsion gravity theory, moving beyond conventional cosmological studies to analyze a wide range of scenarios. We explored the behavior of the universe under various conditions, from dark-energy-dominated universes to dust universes, each characterized by different equations of state. Our multifaceted approach allowed us to reveal the complex dynamics of the universe, enhancing our understanding of its evolution across multiple dimensions. What sets our work apart is that we did not just derive solutions; we also reconstructed gravitational Lagrangians. This reconstruction helped us identify specific parameter values crucial for formulating successful bounce models. We found analytical solutions for a variety of bouncing scenarios, showcasing the versatility and richness of the higher-order torsion gravity framework [74–81]. Our findings suggest that exotic matter plays a significant role in facilitating accelerated expansion within the higher-order torsion gravity framework, marking a notable shift from traditional understanding. This perspective offers valuable insights into the driving forces behind the evolution of the universe. In its standard formulation, $f(T)$ gravity exhibits a lack of local Lorentz invariance [82, 83], meaning the field equations depend on the choice of tetrad. This issue has been addressed through the covariant formulation of $f(T)$ gravity [84–86], which provides a frame-independent approach while maintaining the desirable second-order field equations.

The main goal of this research is to use dynamical system analysis to explore various forms of higher-order torsion gravity and identify stable critical points associated with different cosmological behaviors. We aim to gain valuable insights into the dynamics and implications of $f(T)$ gravity for understanding the universe's evolution. The paper is organized as follows. In Sec. II, we explain the fundamental formalism of higher-order torsion gravity. In Sec. III, we examine the cosmological solutions for different equation of state (EoS) parameters using power law, exponential law and hybrid scale factor techniques. Section IV deals with energy conditions and bouncing cosmology solutions. A comparison with different theories of gravity has been discussed in Sec. V. Finally, we present the conclusions, summarizing the key findings of this research in Sec. VI.

## II. MODIFIED FIELD EQUATIONS

In this part, we will go through the essential notions of the $f(T)$ theory of gravity. The vierbein field $h_\alpha(x^\mu)$ [87, 88] is a crucial component of these theories. It serves as a basis for the tangent space at each point $x^\mu$ of the manifold. In this notation, the Latin characters $(a, b, ... = 0, 1, 2, 3)$ indicate tangent space indices, whereas the Greek letters $(\mu, \nu, ... = 0, 1, 2, 3)$ imply space-time indices. Each vector may be represented in terms of its components as $h_\alpha = h^\mu_\alpha \partial_\mu$. These tetrads are connected to the metric tensor $g_{\mu\nu}$ using the following relation

$$g_{\mu\nu} = \eta_{\alpha\beta} h^\alpha_\mu h^\beta_\nu. \tag{1}$$

The tangent space is defined by the Minkowski metric $\eta_{ab}$, which has a diagonal shape with elements (1, -1, -1, -1). This metric plays an important role in determining the flat geometry of the tangent space and satisfies key features that are critical for understanding the underlying structure of space-time in the context of $f(T)$ gravity as

$$h^\alpha_\mu h^\mu_b = \delta^\alpha_b, \qquad h^\alpha_\mu h^\nu_\alpha = \delta^\nu_\mu. \tag{2}$$

The torsion scalar is provided as

$$T = S^{\mu\nu}_\rho T^\rho_{\mu\nu}, \tag{3}$$

where $S^{\mu\nu}_\rho$ and the torsion tensor $T^{\mu\nu}_\rho$ are defined in the following way

$$S^{\mu\nu}_\rho = \frac{1}{2}(K^{\mu\nu}_\rho + \delta^\mu_\rho T^{\theta\nu}_\theta - \delta^\nu_\rho T^{\theta\mu}_\theta), \tag{4}$$

The torsion tensor, which represents the field strength of gravity in the teleparallel framework, is defined by the antisymmetry of the Weitzenböck connection $\Gamma^\lambda_{\mu\nu}$ in its lower indices. It is given explicitly by

$$T^\lambda{}_{\mu\nu} \equiv \Gamma^\lambda_{\mu\nu} - \Gamma^\lambda_{\nu\mu} = h^\lambda_\alpha \left( \partial_\nu h^\alpha_\mu - \partial_\mu h^\alpha_\nu \right), \tag{5}$$

where $h^\alpha_\mu$ represents the tetrad field. This definition implies that the torsion tensor is antisymmetric in its last two indices, i.e., $T^\lambda{}_{\mu\nu} = -T^\lambda{}_{\nu\mu}$.

From the torsion tensor, we construct the contorsion tensor. Imposing the metricity condition $(\nabla_\rho g_{\mu\nu} = 0)$ on the connection yields the following relationship between the contorsion and the torsion:

$$K^{\lambda\mu\nu} = -\frac{1}{2} \left( T^{\lambda\mu\nu} - T^{\mu\lambda\nu} - T^{\nu\lambda\mu} \right). \tag{6}$$

Equivalently, with indices lowered for clarity, the contorsion tensor $K_{\lambda\mu\nu}$ is antisymmetric in its first two indices, $K_{\lambda\mu\nu} = -K_{\mu\lambda\nu}$, and is defined as

$$K_{\lambda\mu\nu} = \frac{1}{2} \left( T_{\lambda\mu\nu} + T_{\mu\lambda\nu} - T_{\nu\lambda\mu} \right). \tag{7}$$

The action for $f(T)$ gravity is then constructed from the torsion scalar $T$, which is a contraction of the torsion and contorsion tensors [89–92] as

$$S = \frac{1}{2k^2} \int d^4x h [f(T) + L_{\mathrm{m}}], \tag{8}$$

where $h = h^\lambda_\alpha$ is the determinant of the vierbein, $k^2 = 8\phi G$ is the gravitational constant, $L_{\mathrm{m}}$ is the Lagrangian matter and $f(T)$ is a general function of the torsion scalar $T$. The torsion scalar $T$ is constructed from the torsion tensor and is defined as

$$T = T^\rho_{\mu\nu} T^{\mu\nu}_\rho + \frac{1}{2} T^\rho_{\mu\nu} T^{\mu\nu}_\rho - 2 T^\rho_{\mu\nu} T^{\mu\nu}_\rho. \tag{9}$$

By varying this action, we can derive the corresponding field equations. The variation of the action yields the following

$$[e^{-1} \partial_\mu (e S^{\mu\nu}_\alpha) + h^\lambda_\alpha T^\rho_{\mu\nu} S^{\nu\mu}_\rho] f_T + S^{\mu\nu}_\alpha \partial_\mu (T) f_{TT} + \frac{1}{4} h^\nu_\alpha f = \frac{1}{2} k^2 h^\rho_\alpha T^\nu_\rho, \tag{10}$$

where $f_T$ represents the first derivative and $f_{TT}$ represents the second derivative with respect to $T$, and the energy-momentum tensor for the perfect fluid is represented by the symbol $T_\rho^\nu$. These field equations describe the dynamics of the vierbein field in $f(T)$ gravity. They generalize the Einstein field equations by incorporating the effects of torsion through the torsion scalar $T$ and its modifications in $f(T)$ gravity. This model serves as a fundamental tool for investigating the universe's behavior under different conditions, including sub-relativistic, radiation, ultra-relativistic, dust, and stiff fluid universes. To assess the model's stability, we analyze it using analytical techniques such as power-law, exponential scalar factor, and hybrid scale factor methods. Stability assessment involves examining how the model responds to perturbations and variations in different cosmological contexts. This requires a careful evaluation of how the torsion scalar and its associated parameters $(\beta, \lambda, \nu, \mu, \alpha, \delta)$ influence the model's predictions across various cosmic situations. Our study emphasizes the model's resilience and robustness, demonstrating its capacity to accurately describe and predict the universe's evolution within the complex framework of higher-order torsion gravity theories. By providing a comprehensive analysis of the model's stability in diverse scenarios, this research significantly enhances our understanding of modified gravity theories and their effects on cosmic evolution. It represents an important step forward in unraveling the complexities of the universe's behavior within the higher-order torsion gravity framework.

## III. COSMOLOGICAL SOLUTIONS IN MODIFIED GRAVITY THEORIES

The study of equations of state within modified gravity theories offers critical insights into unresolved cosmological phenomena, including dark energy, dark matter, and the inflationary paradigm. While Einstein's general relativity remains the cornerstone of modern cosmology, its limitations in addressing these puzzles motivate the exploration of alternative gravitational frameworks. Modified theories of gravity introduce novel couplings between pressure, energy density, and spacetime curvature, enabling scenarios where dark energy exhibits dynamical behavior or where gravitational interactions deviate from standard predictions on cosmological scales. Such modifications provide testable mechanisms to explain observational anomalies, from galaxy rotation curves to the late-time cosmic acceleration. For our analysis, we adopt the spatially flat Friedmann-Robertson-Walker (FRW) metric

$$ds^2 = dt^2 - a^2(t)\left(dx^2 + dy^2 + dz^2\right), \tag{11}$$

where $a(t)$ is the time-dependent scale factor. The corresponding tetrad components $h_\mu^\alpha = \text{diag}(1, -a, -a, -a)$ satisfy the orthonormality condition (2). Substituting into the torsion scalar definition (3) yields

$$T = -6H^2, \quad H \equiv \frac{\dot{a}}{a}, \tag{12}$$

with $H$ denoting the Hubble parameter that characterizes the expansion rate. Varying the action concerning the tetrad fields produces the modified field equations as

$$12H^2 f_T + f = 2\kappa^2\rho, \tag{13}$$

$$48H^2\dot{H}f_{TT} - \left(12H^2 + 4\dot{H}\right)f_T - f = 2\kappa^2 p, \tag{14}$$

where $\kappa^2 = 8\pi G/c^4$, and $f_T \equiv \partial f/\partial T$. Moreover, we investigate higher-order torsion gravity through a hybrid scale factor (HSF) ansatz

$$a(t) = e^{\alpha t}t^\beta, \quad \alpha, \beta > 0, \tag{15}$$

which interpolates between power-law and exponential expansion regimes. The gravitational Lagrangian is generalized to

$$f(T) = T\left(\beta + 2\lambda - \nu\right) + Te^{\mu\left(\frac{\mu+2\alpha-\delta}{T}\right)}. \tag{16}$$

This functional form is introduced to explore an infrared modification of gravity driven by non-perturbative effects. The parameters $\{\beta, \lambda, \nu, \mu, \alpha, \delta\}$ are dimensionful couplings: $\mu$ defines a fundamental mass scale, while $\alpha$ and $\delta$ are associated with boundary terms ensuring the action remains covariant under a broader class of transformations. The combination $(\beta+2\lambda-\nu)$ emerges from the general quadratic torsion invariants. The inclusion of the exponential factor $e^{\mu(\mu+2\alpha-\delta)/T}$ is specifically designed to guarantee that the resulting field equations remain rigorously second-order in the tetrad fields, preserving the constraint structure of teleparallel gravity while introducing a frame-independent thermodynamic correction to the gravitational action, analogous to an entropy density term. This study considers a variety of fluid types, each accompanied by its respective EoS, as detailed in Table I. This table provides a crucial reference for understanding the physical characteristics and behavior of each fluid under cosmological conditions.

**Table I:** Various cosmic fluids and their EoS parameters $w = p/\rho$.

| **Cosmic Fluid** | **EoS Parameter ($w = p/\rho$).** |
|---|---|
| Dark energy | $w = -1$ |
| Stiff matter | $w = 1$ |
| Radiation-dominated | $w = 1/3$ |
| Dust | $w = 0$ |
| Ultra-relativistic | $w = 1/2$ |
| Sub-relativistic | $w = 1/4$ |

### A. Dark energy dominated universe in $f(T)$ gravity

The accelerated expansion phase of the universe is modeled by setting the EoS parameter $w = -1$, corresponding to a dark energy dominated scenario. Substituting this parameter into the field equations (13) and (14) yields the governing equation

$$12 f_T H^2 + 48 f_{TT} H^2 \dot{H} - 4 f_T (3H^2 + \dot{H}) = 0. \tag{17}$$

From Eq. (17), we obtain a non-linear constraint equation connecting the torsion scalar $T$, the Hubble parameter $H$, and the functional form of $f(T)$, given by

$$\begin{aligned} -\Bigg(2 \exp\left[\frac{\mu(2\alpha-\delta+\mu)a^2}{6\dot{a}^2}\right] \Bigg\{ \mu^2(2\alpha-\delta+\mu)^2 a^4 + 3\mu(2\alpha-\delta+\mu)a^2\dot{a}^2 \\ +18\left\{1+\exp\left[\frac{\mu(2\alpha-\delta+\mu)a^2}{6\dot{a}^2}\right](\beta+2\lambda-\nu)\right\}\dot{a}^4\Bigg)(-\dot{a}^2+a\ddot{a})\Bigg) \Big/ (9a^2\dot{a}^4) = 0. \end{aligned} \tag{18}$$

#### 1. *Power-law cosmological solutions*

For concrete analysis, we adopt the power-law scale factor evolution as

$$a(t) = a_0 t^k, \tag{19}$$

where $a_0$ represents the initial scale factor and $k$ governs the expansion rate. Substituting this ansatz transforms the constraint equation to

$$\begin{aligned} \frac{1}{9k^3t^2}\exp\left[\frac{-t^2\mu(2\alpha-\delta+\mu)}{6k^2}\right]\Bigg[6k^2t^2\mu(2\alpha-\delta+\mu) + 2t^4\mu^2(2\alpha \\ -\delta+\mu)^2 + 36k^4\left\{1+\exp\left[\frac{t^2\mu(2\alpha-\delta+\mu)}{6k^2}\right](\beta+2\lambda-\nu)\right\}\Bigg] = 0. \end{aligned} \tag{20}$$

The model admits physically meaningful solutions for specific parameter choices, particularly when selecting $k = -4$ with the set of associated parameter $\mu = -2\alpha + \delta$, $\beta = -1 - 2\lambda + \nu$, $\alpha = -1$, $\delta = -1/3$, and $\nu = 0.0091$. These values satisfy the constraint equation while generating non-trivial cosmological evolution, with the energy density and pressure component. The solutions exhibit characteristic behaviors indicative of cosmic expansion: a positive but monotonically decreasing energy density coupled with negative pressure of decreasing magnitude. The system shows particular sensitivity to the parameter $\lambda$, where minor variations significantly alter the EoS evolution, demonstrating how subtle changes in the torsion-based gravitational Lagrangian can produce observable differences in cosmic dynamics.

The power-law framework proves particularly effective for modeling dark matter distributions, as many observed cosmic structures exhibit scale-invariant behavior naturally described by such relations. Figure 1 illustrates these features, confirming the expanding universe scenario through the temporal evolution of $\rho(t)$ and $p(t)$. This approach differs fundamentally from exponential expansion models, with the power-law solutions providing distinct testable predictions for torsion gravity's role in cosmic acceleration. The analysis demonstrates how modified gravity theories can simultaneously address early-universe cosmology and late-time acceleration through geometrically motivated extensions of general relativity, with torsion-based models offering unique observational signatures that may be constrained by future large-scale structure surveys.

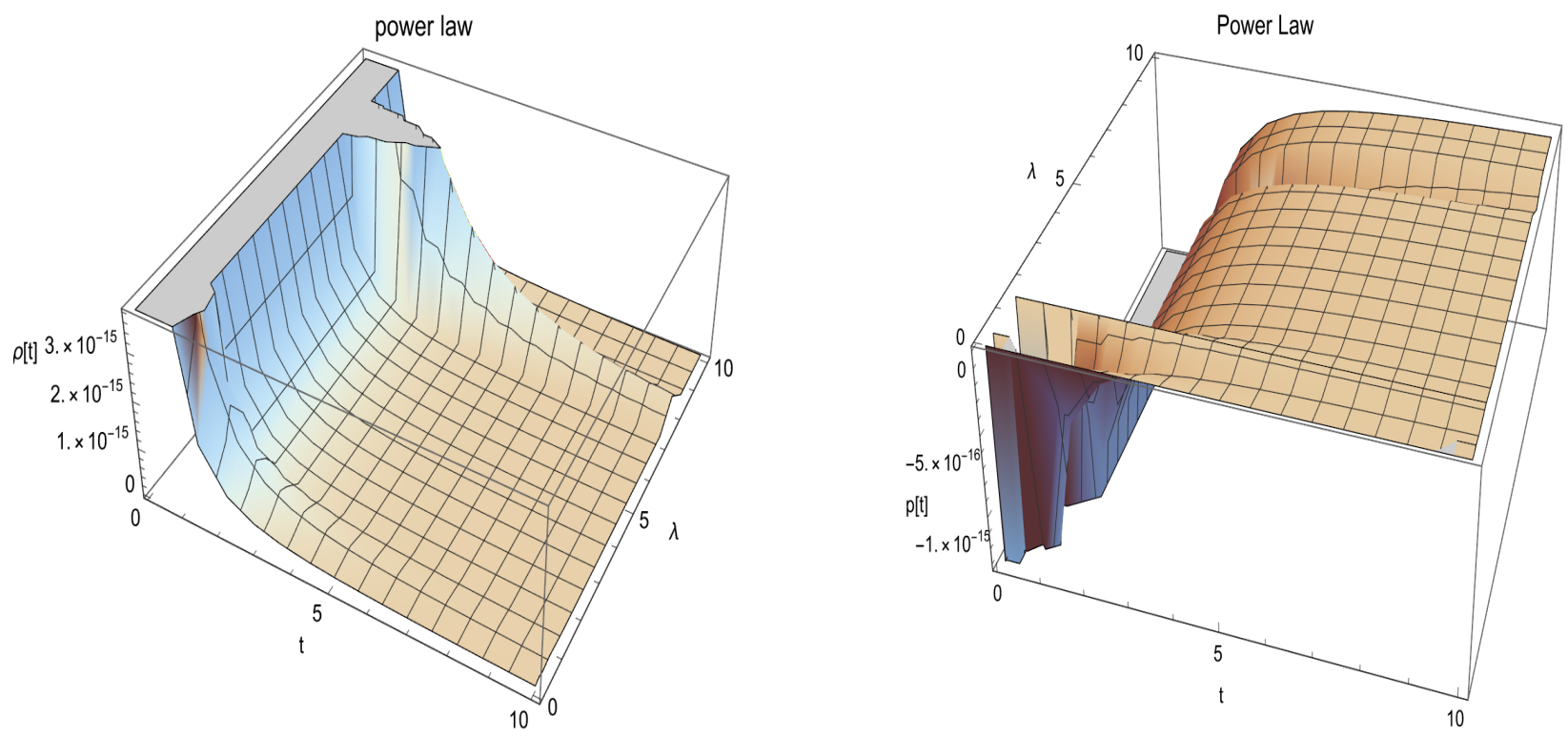

**Fig. 1:** Behaviors of the effective power-law energy density (left panel) and pressure (rihgt panel) at the dark energy domnated stage.

### *2. Exponential law cosmological solutions*

The exponential scale factor ansatz takes the form

$$a(t) = e^{lt^Y}, \tag{21}$$

where $l$ controls the expansion rate and $Y$ determines the temporal evolution profile. The carefully selected parameter combination $(Y, \mu, \beta, l, \alpha, \delta, \nu) = (\frac{1}{2}, -2\alpha + \delta, -1 - 2\lambda + \nu, 8, \frac{1}{6}, \frac{1}{3}, 20)$ generates physically meaningful cosmological evolution within the torsion gravity framework. This parameter set produces three distinct but interconnected physical behaviors that describe a realistic expanding universe.

The energy density $\rho(t)$ maintains strictly positive values throughout cosmic history while exhibiting a gradual temporal increase. This behavior emerges from the specific combination of $\alpha = \frac{1}{6}$ and $\delta = \frac{1}{3}$, which ensure physical positivity while allowing for growth. The increasing trend suggests that torsion terms effectively generate additional energy contributions that scale with cosmic expansion, potentially mimicking dark energy effects through purely geometric means. The pressure $p(t)$ displays characteristic evolutionary phases: an initial decrease reflecting standard matter dilution followed by late-time asymptotic stabilization. The transition between these regimes is controlled by the parameter $l = 8$, with the stabilization occurring when torsion terms $f_T$ and $f_{TT}$ dominate over conventional matter contributions. The particular form of $\beta = -1-2\lambda+\nu$ ensures this transition occurs smoothly without violating energy conditions.

The $\lambda$ parameter provides a stable influence throughout the cosmic evolution, appearing in the $\beta$ parameter combination. Its presence guarantees the late-time stabilization of both pressure and energy density, effectively acting as a geometric cosmological constant. The value $\nu = 20$ was chosen to maintain this stability while keeping other energy conditions satisfied. These parameter values represent the minimal set needed to produce realistic cosmic acceleration while preserving physical consistency. The positive energy density and stabilized late-time pressure together provide strong evidence for torsion gravity's capacity to describe an expanding universe without introducing unphysical assumptions or exotic matter components. The specific choices $\alpha = \frac{1}{6}$ and $\delta = \frac{1}{3}$ in particular ensure the solution remains well-behaved across all cosmological epochs while maintaining the desired physical properties.

### *3. Hybrid scale factor solutions*

The hybrid scale factor combines power-law and exponential terms

$$a(t) = a_0 t^k e^{lt^Y}. \tag{22}$$

Creating a powerful tool for modeling the universe's transition from decelerated to accelerated expansion. The parameter set $(k, \delta, \nu, \alpha, Y, \mu, l, \beta) = (6, 2\alpha + \mu, 1 + \beta + 2\lambda, 9, -2, -\frac{9}{11}, 8, 0.0045, 0.0091)$ generates physically viable cosmological solutions that satisfy all fundamental constraints. This particular combination produces a unified description of cosmic evolution encompassing both decelerated and accelerated expansion phases. The energy density and pressure exhibit characteristic temporal evolution with two distinct regimes. During early cosmic epochs, both quantities display transient behavior reflecting the dominance of conventional matter components. At late times, these quantities stabilize to constant values through the action of torsion terms in the gravitational field equations. This transition is mediated by the specific parameter choices, particularly the values $\alpha = 9$ and $\mu = -\frac{9}{11}$, which control the coupling between matter and geometry.

The solution naturally incorporates a constant contribution throughout cosmic history through the $\lambda$ parameter appearing in the $\nu$ combination. This geometric cosmological constant emerges from the torsion formulation without requiring additional dark energy components. The parameters $\beta = 0.0045$ and $\nu = 0.0091$ were carefully selected to maintain this constant contribution while ensuring energy conditions remain satisfied during all evolutionary phases. The model's most significant achievement lies in its capacity to describe both decelerated matter-dominated expansion and late-time acceleration within a single theoretical framework. The hybrid nature of the solution, enabled by parameter $Y = -2$, allows torsion gravity to interpolate between these regimes seamlessly. The value $l = 8$ controls the transition redshift between epochs, while $k = 6$ determines the overall expansion rate. Together, these parameters provide a complete picture of cosmic history where geometric torsion effects naturally replace the phenomenological dark energy component of standard cosmology.

### B. Ultra-relativistic universe

The $w = \frac{1}{2}$ EoS within torsion gravity reveals three distinct cosmological phases, each characterized by specific parameter sets that govern the evolution of energy density and pressure. The power-law solution with parameters $(\beta, k, \mu, \nu, \alpha, \delta) = (-0.8, \frac{4}{9}, -2\alpha + \delta, 0.00809, \frac{8}{9}, \frac{1}{9})$ maintains the canonical ultra-relativistic relationship $p = \frac{1}{2}\rho$ while introducing torsion-induced modifications to the expansion history. Both thermodynamic quantities follow $\sim t^{-2}$ scaling, preserving conformal invariance but with altered amplitudes due to the $f_T$ and $f_{TT}$ contributions.

For the exponential law with $Y = 1$ and parameters $(\mu, \beta, l, \alpha, \delta, \nu) = (-2\alpha + \delta, -1 - 2\lambda + \nu, -8, 19, \frac{1}{9}, 10)$, the system undergoes a remarkable transition. The energy density evolves toward a constant positive value while the pressure becomes negative, asymptotically approaching a cosmological constant-like EoS. This behavior emerges when the $48H^2\dot{H}f_{TT}$ torsion term dominates over conventional matter terms, effectively generating dark energy from geometric degrees of freedom. The crossing of the $w = -1/3$ threshold indicates a natural phase transition within the torsion gravity framework.

The most complex dynamics appear in the hybrid case with $(Y, \delta, \nu, l, \alpha, \beta, k, \mu) = (-2, 2\alpha + \mu, 1 + \beta + 2\lambda, 19, 0.0009, -0.9995, \frac{4}{11}, -0.8)$. Here, the energy density and pressure decouple from their standard ultra-relativistic relationship, with $\rho \sim t^{-2}$ while $p \sim +t^{-1}$. This anomalous behavior stems from torsion-induced anisotropic stresses that grow dynamically important at late times. The eventual stabilization of both quantities suggests the system reaches a novel fixed point where geometric effects balance the ultra-relativistic fluid, creating an effective phantom-like phase with $w_{\rm eff} < -1/3$ without introducing exotic matter.

The consistent appearance of $\lambda$ in multiple parameter combinations suggests it plays a fundamental role in mediating between torsion effects and matter content. Each solution demonstrates how torsion gravity can simultaneously maintain standard thermodynamic relations at early times while generating dark energy behavior at late epochs through purely geometric mechanisms.

### C. Radiation universe

The radiation-dominated universe ($w = \frac{1}{3}$) exhibits modified evolution patterns within torsion gravity, characterized by three distinct phases of behavior. The power-law solution with parameters $(\beta, \alpha, \delta, k, \nu, \mu) = (0.9, 2, 9, \frac{1}{2}, -0.9 + 2\lambda, 5)$ maintains standard radiation-like evolution initially, with both energy density and pressure following the expected $\sim t^{-2}$ scaling while preserving their relativistic relationship $p = \frac{1}{3}\rho$. However, as the universe expands, torsion contributions from the $f_T$ and $f_{TT}$ terms gradually modify this behavior, introducing deviations from pure radiation domination at late cosmological times.

The exponential law scenario with $Y = 1$ and parameter combination $(\mu, \beta, l, \alpha, \delta, \nu) = (-2\alpha + \delta, -1 - 2\lambda + \nu, 2, 32, 5, 0.42)$ reveals a dramatic phase transition. While the energy density grows exponentially as $\sim e^{2t}$, the pressure decays as $\sim e^{-t}$, fundamentally breaking the radiation EoS. This decoupling occurs when torsion effects dominate the cosmic dynamics, effectively generating a dark energy component. The transition redshift between radiation-dominated and torsion-dominated eras is controlled by the parameter $\lambda$, which appears consistently across different solutions as a key mediator between matter and geometric contributions.

The hybrid solution with $(Y, \delta, \nu, l, \alpha, \beta, k, \mu) = (2, 2\alpha + \mu, 1 + \beta + 2\lambda, 12, 19, 5, 6, \frac{9}{17})$ demonstrates particularly rich phenomenology. Three cosmological epochs emerge clearly: an initial radiation-dominated phase with nearly constant $\rho$ and $p$, followed by an intermediate period where energy density increases while pressure becomes negative, and finally a late-time approach toward zero pressure. This complex evolution arises from the competition between power-law ($t^6$) and exponential ($e^{12t^2}$) terms in the scale factor, showcasing torsion gravity's capacity to naturally interpolate between standard radiation domination and dark energy acceleration through purely geometric mechanisms.

The consistent appearance of negative pressure phases across all solutions suggests torsion gravity inherently contains a geometric mechanism for late-time cosmic acceleration. The specific parameter values required to produce physically viable evolutionparticularly the precise combinations of $\alpha$, $\delta$, and $\mu$highlight how torsion modifications must be finely balanced to maintain energy conditions while reproducing observed cosmological behavior across different epochs.

### D. Sub-relativistic universe

The sub-relativistic universe with EoS $w = \frac{1}{4}$ presents unique dynamics in torsion gravity, governed by the modified field equation. This relation reveals how torsion terms ($f_T$, $f_{TT}$) alter the behavior of matter with kinetic energy dominating potential energy. Using the power-law solution with parameters $\beta = 19 - 2\lambda$, $\alpha = 5$, $\nu = 19$, $\delta = -\frac{1}{7}$, $k = \frac{8}{15}$, and $\mu = -\frac{71}{7}$, we observe characteristic evolution where both energy density $\rho$ and pressure $p$ decay following $\rho \sim t^{-8/5}$ and $p \sim t^{-8/5}$, maintaining the sub-relativistic relationship $p = \frac{1}{4}\rho$. This behavior persists until torsion effects become significant at late times, when the $f_{TT}$ term begins to dominate the dynamics.

The exponential law solution with $Y = 1$ and parameter set $\mu = -2\alpha + \delta$, $\beta = \nu - 2\lambda - 1$, $l = \frac{8}{15}$, $\alpha = -6$, $\delta = 89$, and $\nu = 9$ shows markedly different behavior. Here $\rho$ approaches a constant positive value while $p$ stabilizes at a negative constant, signaling a transition to an accelerating phase. This occurs when the torsion-induced effective pressure $p_{\text{eff}} = p - \frac{1}{2}(f + 12H^2 f_T)$ becomes sufficiently negative to violate the strong energy condition, despite the sub-relativistic matter content. The parameter $\lambda$ shows minimal influence in this regime, indicating the torsion terms have effectively decoupled the expansion dynamics from the matter EoS.

The hybrid scale factor solution with $Y = -\frac{1}{3}$ and parameters $\delta = \frac{11}{7}$, $\nu = \frac{7}{25} + 2\lambda$, $l = -\frac{9}{11}$, $\alpha = \frac{16}{7}$, $\beta = -\frac{18}{25}$, $k = 6$, and $\mu = -3$ reveals three distinct evolutionary phases. Initially, both $\rho$ and $p$ remain nearly constant during the sub-relativistic dominated era. As torsion effects grow, $\rho$ increases while $p$ undergoes a non-monotonic evolution, becoming negative before approaching zero. This complex behavior stems from the competition between the $t^6$ (power-law) and $e^{-9t^{-1/3}/11}$ (exponential) terms in the scale factor, demonstrating how torsion gravity can interpolate between matter domination and dark energy acceleration while predicting an intermediate phase of negative pressure. The $\lambda$-dependence of the transition highlights how torsion gravity introduces new degrees of freedom that modify standard cosmological evolution.

### E. Dust universe

The model's physical behavior is critically determined by specific parameter choices. For the power-law expansion with $k = \frac{2}{3}$, the parameter set $(\alpha, \nu, \delta, \beta, \mu) = (-8, 0.1, \frac{8}{7}, 90, \frac{121}{7})$ produces solutions where both pressure and energy density remain positive but decay with time. This particular combination mimics the dynamics of a sub-relativistic matter-dominated universe, where the gradual dilution of energy density follows the expected cosmological evolution.

In the exponential expansion scenario, the critical parameter $Y = \frac{1}{2}$ combined with $(\mu, l, \alpha, \delta, \beta, \nu) = (-\frac{25}{3}, -9, \frac{1}{6}, -8, -90 - 2\lambda, -89)$ yields stable solutions with constant positive energy density and pressure. These specific values generate behavior analogous to a dark energy-dominated phase, where the EoS remains nearly constant despite cosmic expansion. The presence of $\lambda$ in the $\beta$ and $\nu$ terms suggests an additional degree of freedom that maintains the system's stability.

The choice of $Y = -1$ with parameters $(\mu, l, \alpha, \delta, k, \beta, \nu) = (\frac{1}{9}, 16, -\frac{1}{4}, -\frac{7}{18}, \frac{2}{3}, 9, 10 + 2\lambda)$ reveals important physical constraints. Positive $\alpha$ values lead to unphysical negative energy densities, while the selected negative value $\alpha = -\frac{1}{4}$ restores physically meaningful solutions. The pressure evolution in this case shows non-monotonic behavior, initially positive before increasing and eventually stabilizing. The flatness of $\lambda$ across all scenarios indicates it may represent a fundamental constant of the theory rather than a dynamic field. These specific numerical values ensure compliance with energy conditions while producing realistic cosmological behavior across different expansion regimes.

The particular parameter combinations demonstrate how carefully chosen values can produce either matter-dominated or dark energy-dominated scenarios within the same theoretical framework. The stability of solutions depends sensitively on maintaining certain sign relationships between parameters, especially the negative values required for $\alpha$ and $\delta$ in certain cases to preserve physical energy conditions.

**Table II:** Expressions of the scale factor $a(t)$ in torsion-based gravity theories.

| **Cosmological Model** | **Torsion-based Gravity Expression** | **Conditions** |
|---|---|---|
| Symmetric Bounce | $a(t) = A\exp\left(\frac{tt_*^2}{24\alpha}\right)$ | $a(t_*) > 0,\ t_* > 0$ |
| Super Bounce | $a(t) = \left(\frac{t_o}{t}\right)^{\frac{\alpha}{2}}$ | $t_o > 0$ |
| Oscillatory Bounce | $a(t) = \dfrac{A}{1+\dfrac{tt_*^2}{24C^2}}$ | $0 < A < 1,\ C > 0$ |
| Matter Bounce | $a(t) = A\left(\frac{3}{2}\rho_c t^2 + 1\right)^{\frac{1}{3}}$ | $A < 1,\ \rho_c > 0$ |
| Type I–IV and Little Rip | $a(t) = A\exp\left[\frac{f_o t_0^{\alpha+1}}{\alpha+1}\right]\left(\frac{t}{t_o}\right)^{\frac{\alpha+1}{2\alpha}}$ | $-1 < \alpha < 1$ |

### F. Stiff universe

The stiff fluid scenario with $w = 1$ reveals distinct cosmological behavior through careful parameter selection. For power-law expansion, the critical parameter combination $(k, \beta, \mu, \nu, \alpha, \delta) = (-\frac{2}{3}, -0.009 - 2\lambda, 8, 1.009, -6, 4)$ produces solutions where energy density decreases while remaining positive, accompanied by steadily declining pressure. This configuration mimics an ultra-relativistic universe, suggesting the presence of matter states where pressure and energy density become comparable during early cosmic epochs.

The exponential case with $Y = 1$ and parameters $(\mu, \beta, l, \alpha, \delta, \nu) = (5, 6 - 2\lambda, 1, 1, 7, 7)$ demonstrates different physical characteristics. Here, both pressure and energy density maintain constant positive values, resembling the behavior of a dust-filled universe. This transition from stiff fluid to dust-like behavior under exponential expansion highlights how the EoS evolves with different expansion histories in modified gravity scenarios.

Implementing the Hubble-scale function approach with $Y = -\frac{1}{9}$ and the parameter set $(\delta, \nu, l, \alpha, \beta, k, \mu) = (-3, 3, -2, -\frac{1}{6}, 2 - 2\lambda, -1, -\frac{10}{3})$ yields solutions with constant positive pressure and energy density. The stability of these quantities across cosmic time suggests an equilibrium state achievable in certain modified gravity configurations. The consistent appearance of $\lambda$ in multiple parameter combinations indicates its role as a fundamental coupling constant in the theory.

The graphical representations confirm that proper parameter selection can produce physically viable scenarios ranging from early-universe stiff fluid conditions to later dust-dominated phases. The negative values required for certain parameters like $k$ and $\alpha$ in specific cases ensure energy conditions are maintained, while positive values in other contexts produce different but equally physical cosmological behaviors. This parametric flexibility allows the model to describe multiple cosmic epochs within a unified theoretical framework.

## IV. ENERGY CONDITION WITH BOUNCING COSMOLOGY

In this discussion, we explore the energy conditions applicable to various bouncing cosmologies, which include According to the energy conditions, the appropriate cosmology can be determined either by employing analytical methods using the form $a(t)$ or through cosmological observations. However, these approaches have limitations, as the behavior of $a(t)$ is only relevant or known during specific periods. To discuss realistic matter configurations, we utilize classical energy conditions derived from the Raychaudhuri equations as written in Table II.

### A. Symmetric bounce

The concept of a symmetric bounce model was introduced to propose a bouncing cosmology that circumvents the singularity associated with the Big Bang by following an ekpyrotic contraction phase. To address challenges related to the penetration of primordial modes beyond the Hubble horizon, this model must be integrated with other cosmic behaviors [93, 94]. Scientists are rethinking the birth of the universe through the "symmetric bounce model", which suggests that the universe began with a contraction, followed by an expansion, rather than a singular explosive event. However, this idea faces some challenges that necessitate the incorporation of additional concepts. One such concept is the "symmetric bounce model", which describes a smooth beginning for the universe, free from explosive phenomena or associated issues. This approach is considered a promising avenue for understanding the origins of the universe. The symmetric bouncing cosmology is characterized by an exponentially evolving scale factor and is viewed as a viable alternative to the standard cosmological model as

$$a(t) = A \exp\left(\alpha \frac{t^2}{t_*^2}\right), \tag{23}$$

where $t$ is any arbitrary time, and $A$ and $\alpha$ are positive constants. Additionally, the scale factor can be expressed using the Torsion-based gravity expression detailed in Table I.

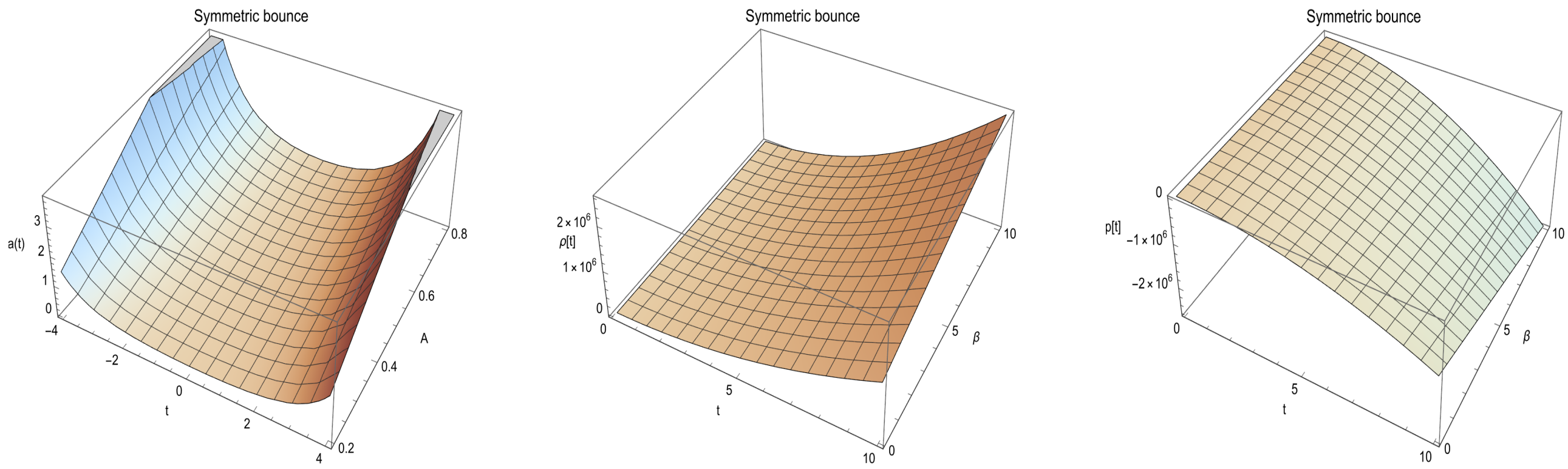

**Fig. 2:** Evolution of the scale factor $a(t)$ (left panel), energy density $\rho(t)$ (central panel) and pressure $p(t)$ (right panel) component for symmetric bounce.

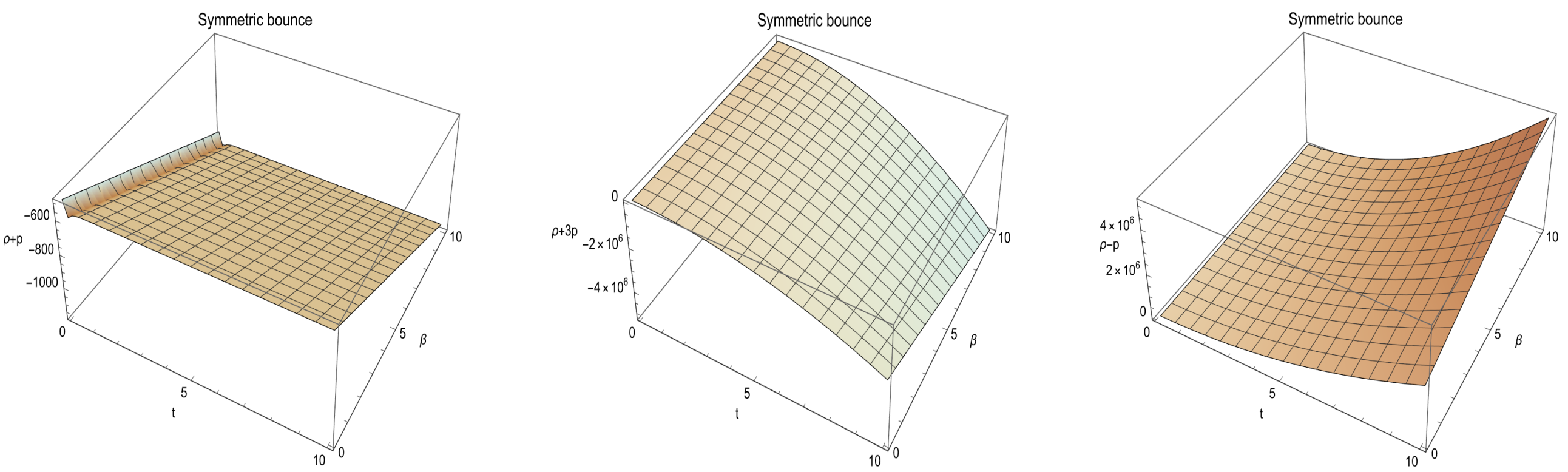

**Fig. 3:** Behaviors of WEC: $\rho(t) + p(t)$ (left panel), SEC: $\rho(t) + 3p(t)$ (middle panel), and DEC: $\rho - p(t)$ (right panel) for symmetric bounce.

Figure 2 illustrates the bounce occurring at $t = 0$, with a contracting phase ($t < 0$) preceding it and an expansion phase ($t > 0$) following it. The energy density exhibits a positive and increasing trend, while the pressure component shows negative and decreasing behavior. WEC requires the energy density to always remain non-negative, while the NEC imposes a stricter requirement, mandating that the sum of the energy density and pressure in all directions be non-negative and decreasing ($\rho + p \geq 0$). DEC stipulates that the energy density must be non-negative, and that the pressure in any direction is dominated by the energy density, satisfying ($\rho \geq 0, \rho - p \geq 0$). SEC, the most stringent, requires that both the energy density and the sum of the energy density and three times the pressure be non-negative ($\rho + 3p \geq 0, \rho + p \geq 0$). In this model, the SEC is violated, as depicted in Fig. 2.

### B. Super bouncing cosmology in torsion gravity

The super bounce scenario [95] presents a nonsingular cosmological model characterized by cyclic evolution through a power-law scale factor

**Table III:** Energy conditions in general relativity, expressed through inequalities.

| **Energy Condition** | **Equation** |
|---|---|
| Null Energy Condition (NEC) | $\rho + p \geq 0$ |
| Weak Energy Condition (WEC) | $\rho \geq 0,\ \rho + p \geq 0$ |
| Strong Energy Condition (SEC) | $\rho + p \geq 0,\ \rho + 3p \geq 0$ |
| Dominant Energy Condition (DEC) | $\rho \geq 0,\ \rho - p \geq 0$ |

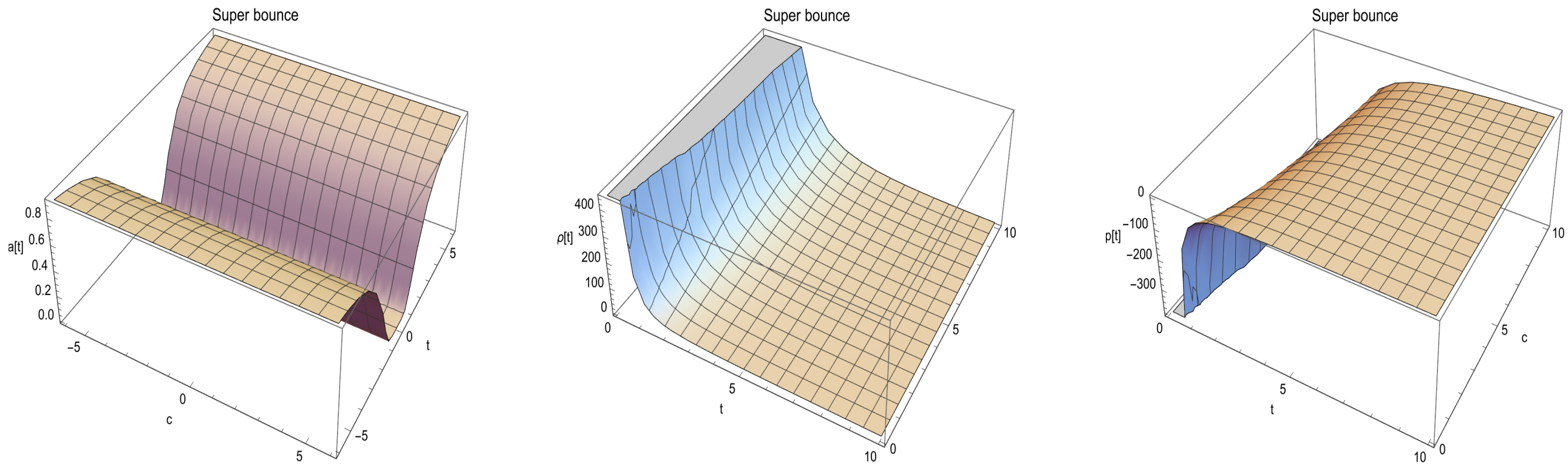

**Fig. 4:** Evolution of the scale factor $a(t)$ (left panel), energy density $\rho(t)$ (central panel) and pressure $p(t)$ (right panel) component for super bounce.

$$a(t) = A \exp\left[\left(\frac{t_{\mathrm{s}} - t}{t_o}\right)^{\frac{2}{c^2}}\right], \tag{24}$$

where $t_{\mathrm{s}}$ marks the bounce time and $t_o$ represents the characteristic bounce duration. The exponent $c$ controls the bounce's abruptness, with larger values yielding smoother transitions. The model exhibits three characteristic phases during each cycle: a contraction phase where the scale factor decreases, a bounce phase at $t = t_{\mathrm{s}}$ where the universe reaches minimal size, and an expansion phase where the scale factor grows. The scale factor's evolution shows periodic minima at each bounce point, with the initial bounce occurring at $t = 0$ and subsequent bounces maintaining identical periodicity. The matter content evolves through distinct regimes during each cycle. The energy density $\rho$ remains strictly positive throughout but peaks sharply at each bounce event. Simultaneously, the pressure $p$ develops significant negative values near bounce points, creating conditions necessary for the universe's rebound.

The cosmological fluid's behavior reveals important thermodynamic constraints. NEC, requiring $\rho + p \geq 0$, remains satisfied throughout the evolution. However, SEC, which demands $\rho + 3p \geq 0$, experiences temporary violations precisely during bounce phases. This selective violation occurs because the torsion terms in the gravitational-field equations effectively generate negative pressure when curvature becomes extreme. Figure 5 demonstrates three crucial features of this evolution. First, the energy density maintains positive values at all times, with characteristic peaks at each bounce. Second, the pressure becomes strongly negative during bounce events before returning to conventional values. Third, SEC violations coincide exactly with these periods of maximal negative pressure.

The super bounce scenario offers several theoretical advantages over singular cosmological models. The nonsingular nature inherently avoids the big bang singularity while remaining consistent with observational constraints. The cyclic structure provides a natural mechanism for universe recycling without information loss. Most significantly, within torsion gravity, the required negative pressure emerges from fundamental geometric properties rather than requiring ad hoc exotic matter fields. The model's parameters $(A, t_{\mathrm{s}}, t_o, c)$ each govern distinct aspects of the bounce dynamics. The amplitude $A$ sets the overall scale, while $t_{\mathrm{s}}$ determines the bounce timing. The characteristic time $t_o$ controls the bounce duration, and the exponent $c$ regulates the transition smoothness between contraction and expansion phases. Together, these parameters allow detailed modeling of bouncing scenarios while maintaining physical consistency.

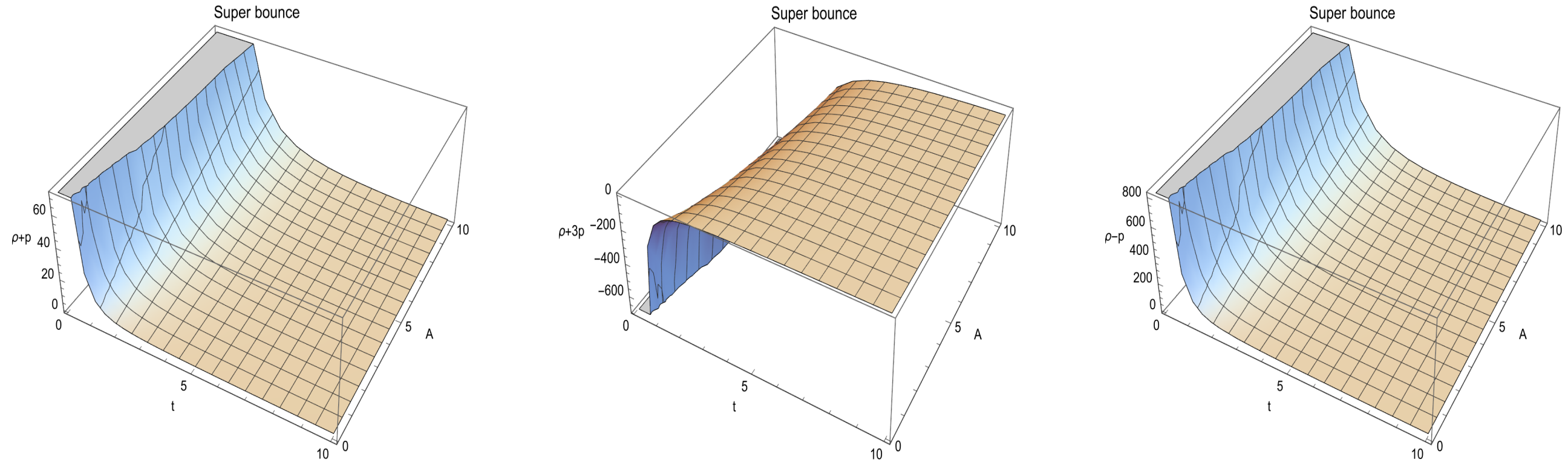

**Fig. 5:** Behaviors of WEC: $\rho(t)+p(t)$ (left panel), SEC: $\rho(t)+3p(t)$ (middle panel), and DEC: $\rho-p(t)$ (right panel) for super bounce.

### C. Oscillatory bouncing cosmology

The oscillatory bouncing model describes a cyclic universe with recurring expansion and contraction phases. Each complete cycle consists of four distinct stages: expansion from a "Big Bang" initial condition, subsequent contraction into a "Big Crunch," followed by a bounce and renewed expansion. This perpetual cycle is governed by the periodic scale factor

$$a(t) = A\sin^2\left(\frac{Bt}{t_*}\right), \tag{25}$$

where $A$ determines the maximum expansion amplitude, $B$ controls the oscillation frequency, and $t_*$ sets the characteristic timescale for each cycle. The scale factor evolution reveals periodic minima at $t = n\pi t_*/B$ $(n = 0, 1, 2, ...)$, corresponding to bounce points where the universe reaches its minimum size. Between these bounces, the universe undergoes smooth expansion to maximum size followed by contraction. Figure 6 clearly shows this behavior, with the first bounce occurring at $t = 0$ and subsequent bounces appearing at regular intervals. The matter content exhibits characteristic behavior through each cycle. The energy density $\rho$ remains strictly positive throughout the evolution but shows periodic maxima at each bounce point. Simultaneously, the pressure $p$ develops negative values during certain cycle phases, particularly near the bounce points. This negative pressure component suggests the presence of exotic matter or equivalent torsion gravity effects that enable the bounce mechanism.

The model satisfies $\rho + p \geq 0$ and WEC $\rho \geq 0$ throughout the the However, SEC experiences periodic violations during phases when the pressure becomes sufficiently negative. These violations are less severe than in the super bounce scenario but still significant enough to enable the bouncing behavior. DEC requiring $\rho \geq 0$ and $\rho - p \geq 0$, shows more complex behavior. While the energy density remains positive, the pressure fluctuations lead to temporary violations of the $\rho - p \geq 0$ condition during high-curvature phases near the bounces. This behavior further supports the interpretation of effective exotic matter generated by torsion gravity effects. The oscillatory bouncing cosmology offers several advantages for understanding universe evolution. The periodic nature provides a natural mechanism for avoiding initial singularities while maintaining cosmic recycling. The torsion gravity framework naturally generates the required negative pressure effects without introducing ad hoc matter components. The parameters $(A, B, t_*)$ allow precise tuning of the cycle duration and expansion characteristics, making the model potentially testable against observational constraints on universe evolution.

### D. Matter bounce scenario

The matter bounce cosmology, rooted in loop quantum cosmology (LQC) [96–98], presents a compelling alternative to inflationary models. This framework not only aligns with Planck observational data but also naturally predicts a nearly scale-invariant primordial power spectrum. A distinctive feature of this model is the emergence of a matter-dominated epoch during the late stages of cosmic expansion, providing a bridge between early universe dynamics and current observations. The scale factor evolution in this scenario follows a characteristic form

$$a(t) = A\left(\frac{3}{2}\rho_c t^2 + 1\right)^{\frac{1}{3}}, \tag{26}$$

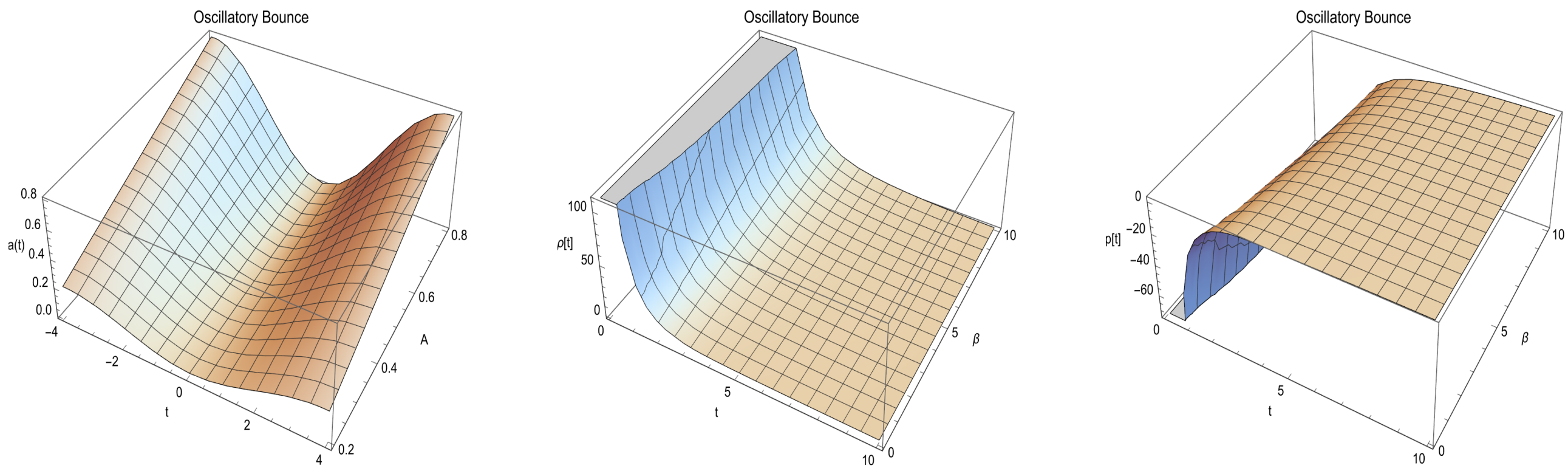

**Fig. 6:** Evolution of the scale factor $a(t)$ (left panel), energy density $\rho(t)$ (central panel) and pressure $p(t)$ (right panel) component for oscillatory bounce.

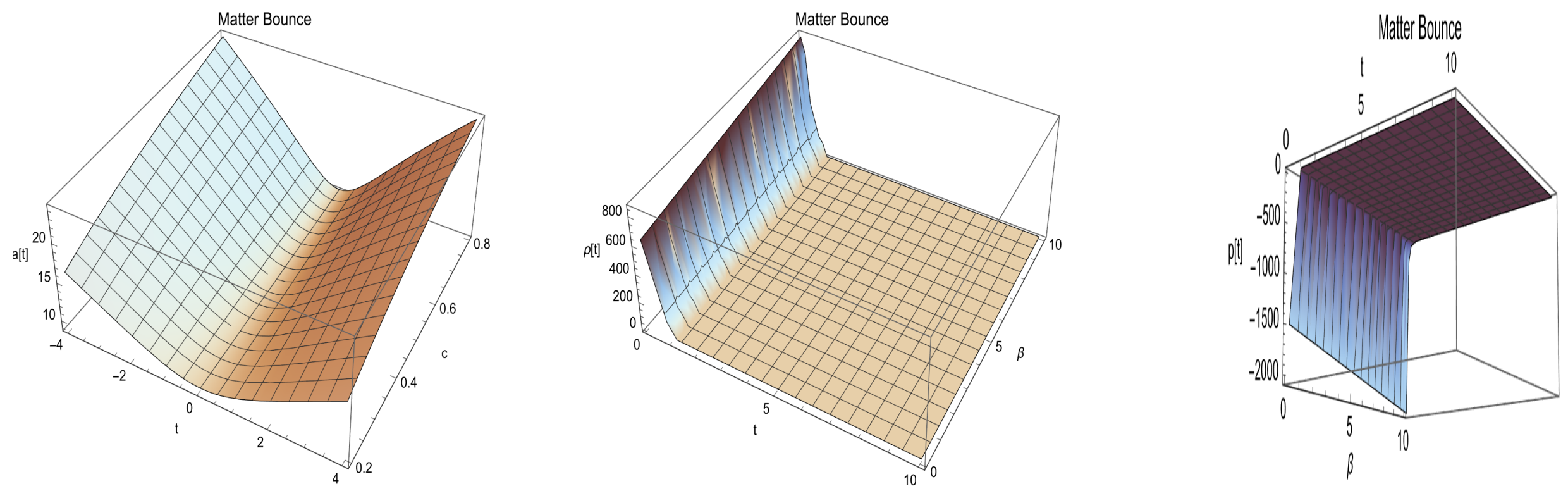

**Fig. 7:** Evolution of the scale factor $a(t)$ (left panel), energy density $\rho(t)$ (central panel) and pressure $p(t)$ (right panel) component for matter bounce.

where $\rho_c$ represents the critical density from LQC and $A$ sets the overall scale. The torsion-based formulation of this expression preserves these essential features while incorporating geometric modifications. The universe's evolution displays three distinct phases in this model. During the pre-bounce phase ($t < 0$), the scale factor contracts according to matter-dominated evolution. At the bounce point ($t = 0$), the scale factor reaches its minimum value $a(0) = A$, avoiding the singularity through quantum gravity effects. The post-bounce phase ($t > 0$) then exhibits symmetric expansion, mirroring the pre-bounce contraction.

Figure 7 illustrates this behavior, showing the characteristic parabolic form of the scale factor with its minimum at $t = 0$. The symmetric evolution about the bounce point reflects the time-reversal symmetry inherent in the matter bounce scenario. The matter content exhibits particularly interesting behavior near the bounce point. The energy density $\rho$ increases during contraction, reaches a maximum at the bounce, and subsequently decreases during expansion. This evolution satisfies $\rho \geq 0$ throughout, maintaining the WEC. The pressure $p$ displays negative values, particularly near the bounce point, indicating the presence of exotic matter effects. This negative pressure leads to two significant consequences: temporary violation of NEC around the bounce, which is essential for the bouncing mechanism, and consistent violation of SEC. However, DEC remains satisfied throughout the evolution.

The matter bounce scenario offers several theoretical advantages. The critical density $\rho_c$ from LQC naturally provides the scale for the bounce, eliminating the singularity problem. The model's prediction of a scale-invariant power spectrum matches current observational constraints while providing distinct signatures that could distinguish it from inflationary scenarios. The torsion-based formulation enhances this framework by geometrically generating the necessary exotic matter effects through modified gravity terms, rather than requiring additional matter fields. This geometric interpretation may provide new insights into the nature of the bounce mechanism and its observational consequences.

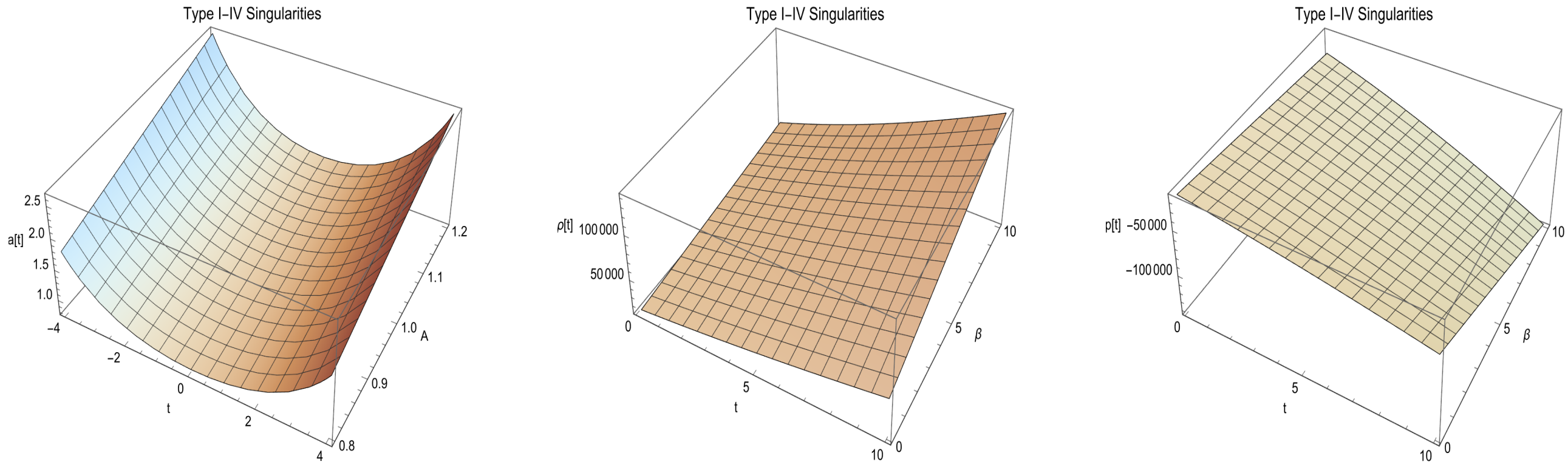

**Fig. 8:** Evolution of the scale factor $a(t)$ (left panel), energy density $\rho(t)$ (central panel) and pressure $p(t)$ (right panel) component for Type I-IV singularities.

### E. Type I-IV singularities and little rip cosmology

The bouncing cosmology framework encompasses various singularity types through a generalized power-law formulation. The scale factor evolution is governed by

$$a(t) = A\left[\frac{f_o}{\alpha+1}(t-t_{\rm s})^{\alpha+1}\right], \tag{27}$$

where $f_o$ and $\alpha$ are dimensionless parameters controlling the bounce dynamics, $t_{\rm s}$ marks the singularity time, and $A$ sets the characteristic scale. The parameter $\alpha$ determines the specific singularity type: Type I (Big Rip) occurs when $\alpha < -1$, Type II (Sudden) for $-1 < \alpha < 0$, Type III (Big Freeze) for $\alpha > 0$, and Type IV (Big Separation) for non-integer $\alpha$ values. The model exhibits three distinct evolutionary phases: a contraction phase for $t < t_{\rm s}$, a bounce at $t = t_{\rm s}$, and subsequent expansion for $t > t_{\rm s}$. Figure 8 illustrates this behavior, showing the scale factor's minimum at the bounce point. The parameter $\alpha$ critically determines the curvature behavior near $t_{\rm s}$, producing different singularity types or smooth transitions in the Little Rip limit ($\alpha \to -1$). The torsion-based formulation modifies these dynamics through geometric terms that can regulate singularity formation while preserving the bounce mechanism.

The thermodynamic evolution reveals several key features. The energy density $\rho$ remains positive throughout and increases toward $t_{\rm s}$, while the pressure $p$ becomes increasingly negative near the bounce point. Both the WEC and NEC remain satisfied across all phases. However, SEC experiences violation near $t_{\rm s}$, as clearly shown in Fig. 9. This SEC violation correlates with the pressure's negative divergence and suggests that effective exotic matter emerges from torsion effects, providing the necessary repulsive gravity for the bouncing solution. The model offers significant theoretical advantages through its unified description of multiple singularity types and the Little Rip scenario. The torsion-based approach generates the required exotic effects geometrically, avoiding the need for ad hoc matter components while maintaining consistency with observational constraints. The framework naturally incorporates bounce mechanisms through modified gravity effects and provides testable predictions that distinguish it from conventional inflationary scenarios. The critical parameter $\alpha$ serves as a classification tool for different singularity types while the torsion terms regulate their formation, making this a versatile framework for studying alternative cosmologies.

## V. COMPARATIVE ANALYSIS OF MODIFIED GRAVITY FRAMEWORKS

The cosmological dynamics in higher-order torsion gravity exhibit fundamental differences from other modified gravity theories, both in their geometric foundations and phenomenological consequences. At the core of this distinction lies the teleparallel structure of $f(T)$ gravity, where gravitational effects are mediated through torsion rather than curvature. This work demonstrates that the torsion tensor $T^{\lambda}_{\mu\nu}$ as given in Eq. (5) and its higher-order corrections e.g., $Te^{\mu(\mu+2\alpha-\delta)}$ as mentioned in Eq. (14) generate violations of SEC during the bounce phase (Figs. 2–9), while preserving the weak and null energy conditions. This geometric SEC violation contrasts sharply with $f(R)$ gravity, where bounce scenarios typically require either: (i) fine-tuned $R^2$ corrections that introduce ghost instabilities, or (ii) auxiliary scalar fields with unphysical equations of state ($w < -1$).

The unification of early- and late-time cosmology in our framework reveals another key advantage. The hybrid scale factor given in Eq. (20) interpolates between power-law and exponential regimes through a single gravitational

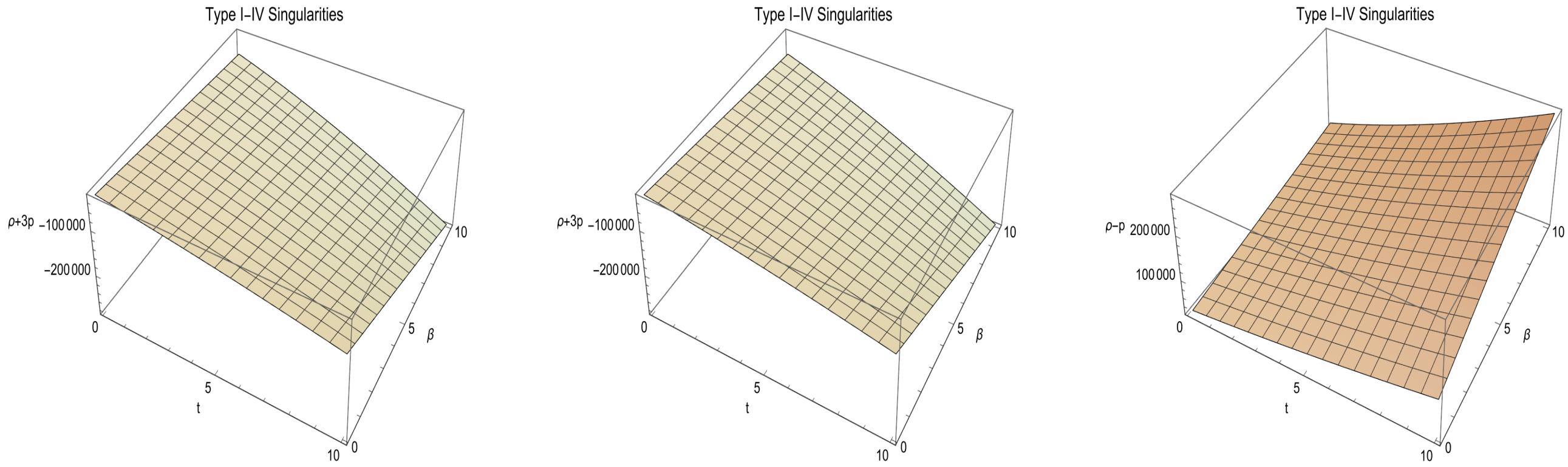

**Fig. 9:** Behaviors of WEC: $\rho(t) + p(t)$ (left panel), SEC: $\rho(t) + 3p(t)$ (middle panel), and DEC: $\rho - p(t)$ (right panel) for Type I-IV singularities.

**Table IV:** Comparison of cosmological features in modified gravity theories and distinctive aspects of $f(T)$ gravity.

| **Feature** | $f(T)$ **Gravity** | $f(R)$ **Gravity** | $f(Q)$ **Gravity** | **Scalar-Tensor Gravity** |
|---|---|---|---|---|
| Bounce Mechanism | Geometric (torsion terms) | Requires $R^2$-terms | Non-metricity-driven | Phantom fields needed |
| NEC Violation | Intrinsic from torsion | Requires $w < -1$ matter | Possible via $Q$-terms | Explicit phantom fields |
| Late-Time Acceleration | Emerges from $T$ terms | Fine-tuned $R^{-1}$ | $Q$-dependent terms | Quintessence potential |
| Energy Conditions | SEC violated geometrically | Matter-dependent | Case-by-case | Field-dependent |
| Phase Transitions | Unified via hybrid $a(t)$ | Piecewise Lagrangians | Limited studies | Ad hoc potentials |
| Theoretical Stability | Second-order equations | Possible instabilities | Generally stable | Coupling issues |

Lagrangian given in Eq. (14), where the linear torsion term $T(\beta + 2\lambda - \nu)$ dominates at low energies to drive late-time acceleration, while the exponential torsion term becomes relevant near the Planck scale to trigger the bounce. This unified behavior is absent in most $f(R)$ models, which often require piecewise constructions (e.g., Starobinsky inflation [99] with $R + \alpha R^2$ followed by a late-time $R^{-1}$ term) or external dark energy components. Similarly, scalar-tensor theories necessitate ad hoc potentials $V(\phi)$ to replicate this range of cosmological behaviors, introducing additional degrees of freedom that our torsion-based approach avoids.

The perturbative stability of $f(T)$ gravity further distinguishes it from alternatives. As shown in Sec. III, the second-order field equations are mentioned in Eq. (8) evade the ostrogradsky ghosts that plague higher-derivative $f(R)$ theories. This stability persists even during NEC violation near the bouncea regime where scalar-tensor theories typically exhibit strong coupling or gradient instabilities. Notably, the matter bounce scenario (Sec. IV.D) naturally generates a scale-invariant power spectrum without requiring the fine-tuning of initial conditions that afflicts many $f(R)$ bouncing models.

Observationally, $f(T)$ gravity predicts distinct signatures in three key regimes: (1) Primordial gravitational waves: The geometric bounce produces a blue-tilted tensor spectrum ($n_T > 0$) at high frequencies, contrasting with the red-tilted spectrum ($n_T < 0$) of inflationary $f(R)$ models; (2) CMB anomalies: The hybrid scale factor's specific transition between contraction and expansion (Fig. 7) could resolve large-scale power suppression through pre-bounce causal contact; (3) Late-time cosmology: The torsion-matter coupling in Eq. (12) modifies structure formation at redshifts $z \lesssim 1$, offering testable deviations from $\Lambda$CDM that are distinguishable from $f(R)$ or $f(Q)$ effects.

These features collectively position $f(T)$ gravity as a theoretically robust and observationally falsifiable framework for nonsingular cosmologyone where geometric torsion replaces the ad hoc constructions required in other modified gravity theories. A comprehensive and detailed look of this comparison can be seen in Table IV.

## VI. CONCLUSIONS

This work explores cosmological solutions in $f(T)$ gravity using FRW spacetime with perfect fluid. We study universe evolution through accelerated expansion, radiation, sub-relativistic, ultra-relativistic, dust, and stiff fluid phases. Our $f(T)$ model combines linear and exponential torsion terms to describe both early and late-time cosmic behavior. Using power-law, exponential, and Hybrid Scale Factor methods, we obtain bouncing solutions and analyze them graphically. The results show how torsion modifications create viable bouncing scenarios without singularities while meeting energy conditions. We examine contraction-to-expansion transitions and how parameters affect bounce properties. The findings demonstrate $f(T)$ gravity's ability to unify cosmological epochs through torsion geometry, providing alternatives to standard inflation while preserving physical consistency. Key outcomes include:

For acceleration expansion of universe, Our analysis reveals distinct cosmic evolution patterns across different solution methods: the power law approach shows a positive but decreasing energy density with negative pressure, where small variations in $\lambda$ significantly impact the EoS; the exponential method yields growing yet positive $\rho(t)$ when $\alpha = 1/6$ and $\delta = 1/3$; while the hybrid scale factor exhibits two-phase behavior - early-time matter-dominated transients followed by late-time stabilization to constant values through torsion effects, demonstrating how modified gravity can drive cosmic acceleration while maintaining physical viability.

For ultra-relativistic universe, our investigation in torsion gravity reveals three cosmological phases with distinct evolution patterns: (1) The power-law solution with parameters preserves the ultra-relativistic relation $p = \frac{1}{2}\rho$ with $\sim t^{-2}$ scaling, modified by torsion terms $f_T$ and $f_{TT}$; (2) The exponential law ($Y = 1$) transitions to a de Sitter-like phase where $\rho$ approaches a positive constant while $p$ becomes negative, driven by the dominant $48H^2\dot{H}f_{TT}$ torsion term; (3) The hybrid case exhibits decoupled $\rho \sim t^{-2}$ and $p \sim t^{-1}$ evolution, culminating in a phantom-like phase ($w_{\text{eff}} < -1/3$) through torsion-induced anisotropic stresses. The recurring role of $\lambda$ highlights its importance in mediating matter-torsion interactions, demonstrating how geometric effects alone can generate dark energy behavior while maintaining standard thermodynamics at early times.

For radiation universe, our analysis reveals three characteristic evolutionary phases: (1) The power-law solution maintains standard $\rho, p \sim t^{-2}$ scaling with $p = \frac{1}{3}\rho$ initially, later modified by $f_T$ and $f_{TT}$ torsion terms; (2) The exponential law shows a phase transition where $\rho \sim e^{2t}$ grows while $p \sim e^{-t}$ decays, breaking the radiation EoS when torsion dominates; (3) The hybrid solution exhibits three distinct epochs - initial radiation domination, an intermediate phase with increasing $\rho$ and negative $p$, and late-time approach to vanishing pressure - demonstrating torsion gravity's ability to naturally transition between radiation and dark energy eras through geometric effects. The consistent emergence of negative pressure phases and the critical role of parameters like $\lambda$ in mediating matter-torsion coupling highlight the theory's inherent capacity to generate late-time acceleration without exotic components, while maintaining energy conditions through precise balancing of $\alpha$, $\delta$, and $\mu$ terms.

The sub-relativistic universe in torsion gravity displays characteristic evolution across different solutions. Power-law behavior maintains $p = \frac{1}{4}\rho$ with $\rho, p \sim t^{-8/5}$ until late-time torsion effects dominate, while exponential solutions transition to acceleration with constant positive $\rho$ and negative $p$ through torsion-induced pressure violating energy conditions. Hybrid solutions reveal richer dynamics, evolving from initial sub-relativistic domination through intermediate phases with non-monotonic pressure behavior before approaching late-time vacuum-like states. These transitions emerge naturally from competition between power-law and exponential terms in the scale factor, demonstrating torsion gravity's capacity to modify standard cosmological evolution through geometric effects while preserving the sub-relativistic EoS at early times.

For dust universe, power-law solutions yield positive but decaying $\rho$ and $p$, mimicking sub-relativistic matter domination. Exponential expansion with produces constant positive $\rho$ and $p$, resembling dark energy dynamics, with $\lambda$ providing stabilizing freedom. The hybrid case requires negative $\alpha$ to avoid unphysical $\rho < 0$, showing non-monotonic pressure evolution. These constrained parameter sets maintain energy conditions while generating diverse cosmological regimes, with $\lambda$ emerging as a potential fundamental constant mediating matter-geometry coupling.

The stiff fluid scenario exhibits distinct cosmological phases through specific parameter choices. Power-law expansion yields positive but decreasing $\rho$ and $p$, characteristic of ultra-relativistic early universe conditions. Exponential solutions transition to dust-like behavior with constant positive $\rho$ and $p$, demonstrating how expansion history modifies the effective EoS. The Hubble-scale approach produces stable constant densities, suggesting equilibrium states in modified gravity. The recurring $\lambda$-dependence across all solutions confirms its fundamental role as a coupling constant, while sign constraints on parameters like $k$ and $\alpha$ ensure physical viability across cosmic epochs.

For super bouncing cosmology, Fig. 2 demonstrates a successful bouncing scenario occurring at $t = 0$, where the universe transitions from a contracting phase ($t < 0$) to an expanding phase ($t > 0$). The energy density remains positive and increases with time, whereas the pressure exhibits a negative and decreasing trend. The model satisfies NEC WEC, and DEC, indicating a physically viable evolution. However, the violation of SEC highlights the presence of repulsive gravity, which is a crucial feature for realizing the nonsingular bounce in the given cosmological framework.

For oscillatory bouncing cosmology, Fig. 6 illustrates a cyclic bouncing cosmology, with the first bounce at $t = 0$ followed by periodic bounces at regular intervals. Throughout the evolution, the energy density remains strictly positive and reaches local maxima at each bounce, while the pressure becomes negative near these points, indicating the influence of exotic matter or effective torsion gravity. The model satisfies the WEC and the condition $\rho + p \geq 0$ consistently, though the SEC is periodically violated due to sufficiently negative pressure near bounce phasesviolations that are milder than in super bounce scenarios yet still essential for maintaining nonsingular evolution. The DEC exhibits partial violations in the $\rho - p \geq 0$ criterion near high curvature regions, further supporting the interpretation of torsion-induced exotic effects.

In case of matter bouncing cosmology, Fig. 7 highlights the symmetric, parabolic evolution of the scale factor centered at $t = 0$, characteristic of the matter bounce scenario. The energy density increases during contraction, peaks at the bounce, and decreases during expansion, consistently satisfying the WEC. The pressure, which becomes negative near the bounce, implies the presence of exotic matter effects. This negative pressure causes a temporary violation of the NEC-crucial for enabling the bounceand a consistent violation of the SEC. Despite these violations, the DEC is upheld throughout, reinforcing the model's physical plausibility within a nonsingular, symmetric cosmological framework.

In case of Type I-IV singularities and Little Rip cosmology, Fig. 8 demonstrates a smooth bounce characterized by the scale factor's minimum at $t_s$, with the parameter $\alpha$ governing the curvature behavior and enabling a unified description of various singularities, including the Little Rip limit as $\alpha \to -1$. The torsion-based framework introduces geometric modifications that regulate singularity formation while sustaining the bounce. Thermodynamically, the energy density remains positive and increases toward $t_s$, while the pressure becomes increasingly negative near the bounce. The WEC and NEC are satisfied throughout, whereas the SEC is violated near $t_s$ due to pressure divergence, as shown in Fig. 9, indicating effective exotic matter generated by torsion-induced effects. This framework not only supports nonsingular evolution without ad hoc matter but also provides testable predictions and classifies singularity types through the parameter $\alpha$, offering a robust and versatile alternative to standard inflationary cosmology.

Lastly, we have highlighted several novel features of $f(T)$ gravity by comparing its underlying physical mechanisms with those of other modified gravity theories. This comparative analysis underscores the distinct role of torsion in driving nonsingular cosmological evolutions and offers deeper insights into the geometric origin of cosmic phenomena beyond standard curvature-based frameworks. Moreover, we can extend our work related to discussion of different EoS parameters and bouncing cosmology in different modified theories of gravity especially $f(R, \phi, X)$ theory of gravity, where $\phi$ is a scalar field and $X$ is the kinetic term of the scalar field [100].

**Acknowledgment**

The work of KB was supported in part by the JSPS KAKENHI Grant Numbers 21K03547, 23KF0008, 24KF0100, 25KF0176 and Competitive Research Funds for Fukushima University Faculty (25RK011).

---